\documentstyle[12pt,a4,psfig]{article}
\catcode`\@=11
\def\marginnote#1{}
\def\ifmath#1{\relax\ifmmode #1\else $#1$\fi}

\def\mstop{m_{\,\widetilde{t}}}
\def\stop{\,\widetilde{t}}

\def\bold#1{\setbox0=\hbox{$#1$}%
     \kern-.025em\copy0\kern-\wd0
     \kern.05em\copy0\kern-\wd0
     \kern-.025em\raise.0433em\box0 }

\def\GENITEM#1;#2{\par\vskip6pt \hangafter=0 \hangindent=#1
   \Textindent{$ #2$ }\ignorespaces}

\newcount\hour
\newcount\minute
\newtoks\amorpm
\hour=\time\divide\hour by60
\minute=\time{\multiply\hour by60 \global\advance\minute by-
\hour}
\edef\standardtime{{\ifnum\hour<12 \global\amorpm={am}%
    \else\global\amorpm={pm}\advance\hour by-12 \fi
    \ifnum\hour=0 \hour=12 \fi
    \number\hour:\ifnum\minute<100\fi\number\minute\the\amorpm}}
\edef\militarytime{\number\hour:\ifnum\minute<100\fi\number\minute}
\def\draftlabel#1{{\@bsphack\if@filesw {\let\thepage\relax
  \xdef\@gtempa{\write\@auxout{\string
    \newlabel{#1}{{\@currentlabel}{\thepage}}}}}\@gtempa
    \if@nobreak \ifvmode\nobreak\fi\fi\fi\@esphack}
     \gdef\@eqnlabel{#1}}
\def\@eqnlabel{}
\def\@vacuum{}
\def\draftmarginnote#1{\marginpar{\raggedright\scriptsize\tt#1}}
\def\draft{\oddsidemargin -.5truein
        \def\@oddfoot{\sl preliminary draft \hfil
        \rm\thepage\hfil\sl\today\quad\militarytime}
        \let\@evenfoot\@oddfoot \overfullrule 3pt
        \let\label=\draftlabel
        \let\marginnote=\draftmarginnote

\def\@eqnnum{(\theequation)\rlap{\kern\marginparsep\tt\@eqnlabel}%
\global\let\@eqnlabel\@vacuum}  }
\def\preprint{\twocolumn\sloppy\flushbottom\parindent 1em
        \leftmargini 2em\leftmarginv .5em\leftmarginvi .5em
        \oddsidemargin -.5in    \evensidemargin -.5in
        \columnsep 15mm \footheight 0pt
        \textwidth 250mmin      \topmargin  -.4in
        \headheight 12pt \topskip .4in
        \textheight 175mm
        \footskip 0pt

\def\@oddhead{\thepage\hfil\addtocounter{page}{1}\thepage}
        \let\@evenhead\@oddhead \def\@oddfoot{} \def\@evenfoot{}
}
\def\titlepage{\@restonecolfalse\if@twocolumn\@restonecoltrue\o
necolumn
     \else \newpage \fi \thispagestyle{empty}\c@page\z@
        \def\thefootnote{\fnsymbol{footnote}} }
\def\endtitlepage{\if@restonecol\twocolumn \else  \fi
        \def\thefootnote{\arabic{footnote}}
        \setcounter{footnote}{0}}  %\c@footnote\z@ }
\catcode`@=12
\relax
\def\be{\begin{equation}}
\def\ee{\end{equation}}
\def\bea{\begin{eqnarray}}
\def\eea{\end{eqnarray}}
\def\simlt{\stackrel{<}{{}_\sim}}
\def\simgt{\stackrel{>}{{}_\sim}}

\def\NPB#1#2#3{{\it Nucl.~Phys.} {\bf{B#1}} (19#2) #3}
\def\PLB#1#2#3{{\it Phys.~Lett.} {\bf{B#1}} (19#2) #3}
\def\PRD#1#2#3{{\it Phys.~Rev.} {\bf{D#1}} (19#2) #3}
\def\PRL#1#2#3{{\it Phys.~Rev.~Lett.} {\bf{#1}} (19#2) #3}

\def\MPLA#1#2#3{{\it Mod.~Phys.~Lett.} {\bf#1} (19#2) #3}

\def\AP#1#2#3{{\it Ann.~Phys.} {\bf#1} (19#2) #3}

\def\HPA#1#2#3{{\it Helv.~Phys.~Acta} {\bf#1} (19#2) #3}
\def\JETPL#1#2#3{{\it JETP~Lett.} {\bf#1} (19#2) #3}

\def\mst11{m_{\;\widetilde{t}_{1}}}

\def\mst22{m_{\;\widetilde{t}_{2}}}
\def\mst12{m_{\;\widetilde{t}_{1,2}}}

\def\msb11{m_{\;\widetilde{b}_{1}}}
\def\msb22{m_{\;\widetilde{b}_{2}}}
\def\msb12{m_{\;\widetilde{b}_{1,2}}}

\def\mwidetilde2{\widetilde{m}^{2}}

\relax
\begin{document}
\topmargin-2.5cm
%\draft
%\preprint
%
\begin{titlepage}
\begin{flushright}
CERN-TH/96-242\\
IEM-FT-140/96 \\
FERMILAB-Pub-96/271-A\\
TURKU-FL-P23-96\\
hep--ph/9702409 \\
\end{flushright}
\vskip 0.3in
\begin{center}
{\Large\bf Electroweak Baryogenesis and Low Energy
Supersymmetry}
\footnote{Work supported in part by the European Union
(contract CHRX/CT92-0004) and CICYT of Spain
(contract AEN95-0195).}
\vskip .5in
{\bf M. Carena~$^{\S}$},
{\bf M. Quir\'os~$^{\ddagger}$},
{\bf A. Riotto~$^{\S}$},
{\bf I. Vilja~$^{\P}$}
and {\bf C.E.M. Wagner~$^{\dagger}$}
\vskip.35in
$^{\S}$FERMILAB, Batavia, IL
60510-0500, USA\\
$^{\dagger}$~CERN, TH Division, CH--1211 Geneva 23, Switzerland\\
$^{\ddagger}$~Instituto de Estructura de la Materia, CSIC, Serrano
123, 28006 Madrid, Spain\\
$^{\P}$~Department of Physics, University of Turku, FIN-20014
Turku, Finland
\end{center}
\vskip1.3cm
\begin{center}
{\bf Abstract}
\end{center}
\begin{quote}
Electroweak baryogenesis is an interesting theoretical scenario,
which demands physics beyond the Standard Model at
energy scales of the order of the weak boson masses.
It has been recently emphasized that, in the
presence of light stops, the electroweak phase transition can be
strongly first order, opening the window for electroweak
baryogenesis in the MSSM. For the realization of this scenario,
the Higgs boson must be light, at the reach of the
LEP2 collider. In this article, we compute the baryon asymmetry
assuming the presence of non-trivial CP violating phases in the parameters
associated with the left-right  stop mixing term and the Higgsino mass $\mu$.
We conclude that a phase $|\sin \phi_{\mu}| > 0.01$ and Higgsino and
gaugino mass parameters $|\mu| \simeq M_2$, and
of the order of the electroweak scale,
are necessary in order to generate the observed baryon asymmetry.
\end{quote}
\vskip1.cm
\begin{flushleft}
%CERN-TH/96-242\\
February 1997 \\
\end{flushleft}

\end{titlepage}
\setcounter{footnote}{0}
\setcounter{page}{0}
\newpage
%
% BODY
\noindent
{\bf 1.} The origin of the observed baryon asymmetry is one of the
most
fundamental open questions in particle physics.
The Standard Model (SM) fulfills all the
requirements~\cite{baryogenesis} for a successful generation
of baryon number~\cite{reviews}, due to the presence of
anomalous processes~\cite{anomaly}, which also
put relevant constraints in models in which the baryon asymmetry
is generated at a very high energy scale~\cite{sphalerons}.
However, the electroweak phase transition is too weakly first order
to assure the preservation of the generated baryon asymmetry at
the electroweak phase transition~\cite{first},
as perturbative~\cite{improvement,twoloop} and
non-perturbative~\cite{nonpert} analyses have shown. On the other
hand,
CP-violating processes are suppressed by powers of $m_f/M_W$,
where $m_f$ are the light-quark masses.
These suppression factors are sufficiently strong
to severely restrict the possible baryon number
generation~\cite{fs,huet}.
Therefore,
if the baryon asymmetry is generated at the electroweak phase
transition,
it will require the presence of new physics at the electroweak scale.

Low energy supersymmetry is a well motivated possibility, and it is
hence highly interesting to test under which conditions there is
room for electroweak baryogenesis in this
scenario~\cite{early,mariano1,mariano2}.
It was recently shown~\cite{CQW}
that the phase transition can be sufficiently strongly first
order only in a restricted region of parameter space: The lightest
stop
must be lighter than the top quark, the ratio of vacuum expectation
values $\tan\beta < 3$, while the lightest Higgs must be at the
reach of
LEP2~\footnote{Explicit two-loop calculations have the general
tendency to
strengthen the phase transition~\cite{CEQW,JoseR} thus making the
previous
bounds as very conservative ones.}.  
Similar results were independently obtained by
the authors of Ref.~\cite{Delepine}.
These results have been confirmed by explicit sphaleron
calculations
in the Minimal Supersymmetric Standard Model
(MSSM)~\cite{MOQ}. 

On the other hand, the Minimal Supersymmetric Standard Model
contains, on top of the Cabbibo-Kobayashi-Maskawa (CKM)
matrix phase, additional sources of CP-violation
and can account for the observed baryon asymmetry.
New CP-violating phases can arise
from the soft supersymmetry breaking parameters associated with
the  stop mixing angle.  Large values of the mixing angle
are, however, strongly restricted in order to preserve a
sufficiently strong first
order electroweak phase
transition~\cite{early,mariano1,mariano2,CQW}.
Therefore, an acceptable baryon asymmetry
may only be generated through a delicate balance between the
values
of the different soft supersymmetry breaking parameters
contributing
to the stop mixing parameter, and their associated CP-violating
phases.
The value of the Higgsino and gaugino masses also play a very
relevant
role in the determination of the baryon asymmetry. Indeed, the
final
baryon asymmetry depends strongly on the relative value of
the Higgsino and gaugino mass parameters with respect to the
critical
temperature  $T_c = {\cal{O}}$(100 GeV).

In this letter we compute the baryon asymmetry and the strength of
the first order phase transition in the MSSM. We identify the region
in the supersymmetric parameter space where baryon asymmetry is
consistent with the observed value $n_B/s \sim 10^{-10}$ and,
furthermore, it is not washed out inside the bubbles after the
phase transition. We obtain that the second effect is guaranteed
provided the light stop mass is in the range
$M_Z \simlt m_{\widetilde{t}} \simlt m_t$,
the lightest
Higgs boson mass is bounded by $m_H\simlt  80$ GeV and the
CP-odd boson has a mass $m_A\simgt 150$ GeV, independently of the
chargino and neutralino masses.
On the other hand baryon asymmetry is generated mainly
by charginos and neutralinos which are not much heavier than the
critical temperature, independently of the mass of the lightest stop.

\vspace{1cm}
\noindent
{\bf 2.} As discussed above, a strongly first order electroweak
phase transition
can  be achieved in the presence of a top squark
lighter than the top quark~\cite{CQW}.
In order to naturally
suppress its contribution to the parameter $\Delta\rho$ and hence
preserve a good agreement with the precision measurements at LEP,
it should be mainly right handed. This can be achieved if the left
handed stop soft supersymmetry breaking mass $m_Q$
is much larger than $M_Z$.
For moderate mixing,
the lightest stop mass is then approximately given by
\be
\mstop^2  \simeq m_U^2 + D_R^2 + m_t^2(\phi) \left( 1  -
\frac{\left|\widetilde{A}_t\right|^2}{m_Q^2}
\right)
\ee
where $\widetilde{A}_t = A_t - \mu^{\star}/\tan\beta$ is the
particular combination appearing in the off-diagonal terms of
the left-right stop squared mass matrix and $m_U^2$ is
the soft supersymmetry breaking squared mass parameter
of the right handed stop.  Observe that the stop mass
eigenvalues and hence, the stop contribution to the Higgs
effective potential must be computed by taking into account the
non-vanishing complex phases for the parameters $A_t$ and
$\mu$.
For most practical purposes, however, unless the phases are of
order one, we can
identify the absolute value of the parameters $A_t$ and $\mu$
with their real components and hence, we shall not distinguish
them in our notation.

We shall work within the framework of the improved one-loop
effective potential. 
Hence two loop corrections~\cite{CEQW,JoseR}
will not be included in our analysis. As stated above,
since these corrections tend to increase
the strength of the first order phase
transition~\cite{JoseR}, our bounds
should be taken as conservative ones.
The preservation of the baryon number asymmetry requires  the
order parameter $v(T_c)/T_c$ to be larger than one, where
$v$ denotes the Higgs vacuum expectation value (VEV) and we have
normalized it such that $v(0) = v = 246.22$ GeV.
Following the analysis of Refs.~\cite{mariano1,CQW},
the order parameter $v(T_c)/T_c$
is bounded to be below the maximum value obtained for
$m_A \gg T_c$,
\be
\frac{v(T_c)}{T_c} < \left(\frac{v(T_c)}{T_c}\right)_{\rm SM}
+ \frac{2 \; m_t^3  \left(1 -
\widetilde{A}_t^2/m_Q^2\right)^{3/2}}{ \pi \; v \; m_H^2}\ ,
\label{totalE}
\ee
where $m_t = \overline{m}_t(m_t)$ is the on-shell running top
quark
mass in the $\overline{{\rm MS}}$ scheme.
The first term on the right hand side of
Eq.~(\ref{totalE}) is the Standard Model contribution
\be
\left(\frac{v}{T}\right)_{\rm SM}
\simeq \left(\frac{40}{m_H[{\rm GeV}]}\right)^2,
\ee
and the second term is
the contribution that would be obtained if the right handed
stop plasma mass vanished at the critical
temperature (see Eq.~(\ref{plasm})).
The lower the value of this mass at the critical temperature,
the larger the value of $v(T_c)/T_c$~\cite{CQW}.
For $m_A \simgt 150$ GeV, preferred to enhance the strength of
the phase transition~\cite{mariano1},
the present experimental bound on the Higgs mass reads
$m_H \geq 65\ {\rm GeV}$,
meaning that the SM contribution leads to a value of $v(T_c)/T_c$
at most of order $1/3$.

In order to overcome the Standard Model
constraints, the stop contribution must be large.
The stop contribution strongly depends
on the value of $m_U^2$, which must be small in magnitude, and
negative, in order to induce a sufficiently strong first order phase
transition. Indeed, large stop contributions
are always associated with small values of the right handed stop
plasma mass
\begin{equation}
m^{\rm eff}_{\;\widetilde{t}} = -\widetilde{m}_U^2 + \Pi_R(T)
\label{plasm}
\end{equation}
where $\widetilde{m}_U^2 = - m_U^2$, $\Pi_R(T) \simeq 4 g_3^2
T^2/9+h_t^2/6[2-
\widetilde{A}_t^2/m_Q^2]T^2$~\cite{CQW,CE} is the finite
temperature
self-energy contribution to the right-handed
squarks~\footnote{We are considering heavy (decoupled from the
thermal bath) gluinos. For light gluinos, their contribution
to the squark self-energies,
$2g_3^2 T^2/9$, should be added to $\Pi_R(T)$~\cite{mariano1}.}
and $h_t$ and $g_3$  are the
top quark Yukawa and strong gauge couplings, respectively.
Moreover, the
trilinear mass term, $\widetilde{A}_t$,
must be $\widetilde{A}_t^2 \ll m_Q^2$
in order to avoid the suppression of  the stop contribution
to $v(T_c)/T_c$.

Although large values of $\widetilde{m}_U$, of order of the critical
temperature,
are useful to get a strongly first order phase transition, they may
also induce charge and color breaking minima. Indeed, if the
effective plasma mass at the critical temperature vanished, the
universe would be driven to a charge and color breaking minimum at
$T \geq T_c$~\cite{CQW}. Hence, the upper bound on $v(T_c)/T_c$,
Eq.~(\ref{totalE}) cannot be reached in realistic scenarios. A
conservative
bound on $\widetilde{m}_U$ may be obtained by demanding that
the electroweak symmetry breaking minimum should be lower
than any
color-breaking minima induced by the presence of
$\widetilde{m}_U$ at
zero temperature, which yields the condition
\begin{equation}
\widetilde{m}_U \leq \left(\frac{m_H^2 v^2 g_3^2}{12}\right)^{1/4}.
\label{colorbound}
\end{equation}
It can be shown that this condition is sufficient to prevent dangerous
color breaking minima at zero and finite temperature for any value of
the mixing parameter $\widetilde{A}_t$ \cite{CQW}.
In this work, we shall use
this conservative bound.

In order to obtain values of $v(T_c)/T_c$ larger than one,
the Higgs mass must take  small values, close to the present
experimental bound.  Numerically,
an upper bound, of order 80 GeV, can be derived.
For small mixing, the one-loop Higgs mass has a very simple form
\begin{equation}
m_H^2 = M_Z^2 \cos^2 2\beta + \frac{3}{4\pi^2}
\frac{\overline{m}_t^4}{v^2}
\log\left(\frac{m_{\widetilde{t}}^2 m_{\widetilde{T}}^2}
{\overline{m}_t^4}\right)\left[1
+ {\cal{O}}\left(\frac{\widetilde{A}_t^2}{m_Q^2}\right)
\right],
\end{equation}
where $m_{\widetilde{T}}^2 \simeq m_Q^2 + m_t^2$, is the heaviest
stop
squared mass. Hence, $\tan\beta$ must take values close to one. The
larger the left handed stop mass, the closer to one $\tan\beta$
must be.
In this work we shall take $m_Q \geq 500$ GeV. This implies that
the left handed stop effects decouple at the critical temperature
and hence,
different values of $m_Q$ mainly affect the baryon asymmetry through
the resulting Higgs mass.

\vspace{1cm}
\noindent
{\bf 3.} Concerning the baryon asymmetry generation,
the new source of CP-violation, beyond the one contained
in the CKM matrix,
may be either explicit~\cite{ex1} or
spontaneous~\cite{noi} in the Higgs sector (which requires
at least two Higgs doublets). In both cases,
particle mass matrices acquire a
nontrivial space-time dependence when bubbles of the broken
phase
nucleate and expand during a first-order electroweak phase
transition. The crucial observation is that this
space-time dependence cannot be rotated away at two
adjacent points by the same unitary transformation. This provides
sufficiently
fast nonequilibrium CP-violating  effects inside the wall of a
bubble of broken phase expanding in the plasma and may give
rise to a nonvanishing baryon asymmetry through the anomalous
$(B+L)$-violating transitions~\cite{sp} when particles
diffuse to the exterior of the advancing bubble.

Baryogenesis is therefore
fueled by CP-violating sources which  are
locally induced by the passage of the wall~\cite{thick,thicknoi}.
These sources should
be inserted into a set of classical Boltzmann
equations describing  particle
distribution densities and permitting to take into account
Debye screening of induced
gauge charges~\cite{deb}~\footnote{This effect will be ignored
since the impact on the final
result is ${\cal O}(1)$~\cite{deb}.},
particle number changing reactions~\cite{cha} and
to trace the crucial role
played by diffusion~\cite{tra}.
Indeed,
transport effects  allow  CP-violating charges to  efficiently
diffuse in
front of the advancing bubble wall where anomalous electroweak
baryon
violating processes are unsuppressed.
This amounts to  greatly enhancing the final baryon asymmetry.

We shall make use of a method  recently proposed by one of
the authors~\cite{riotto} to compute the
effect of CP-violation coming from extensions of the Standard Model
on the
mechanism of electroweak baryogenesis. This method  is entirely
based on
a nonequilibrium quantum field theory
diagrammatic approach and has two main virtues: it   may be
applied
for all wall shapes and sizes of the bubble wall and it naturally
incorporates the effects of the  incoherent nature of plasma physics
on
CP-violating observables.
The method relies on the closed-time path (CPT) formalism,
which is a powerful Green's function
formulation for describing nonequilibrium phenomena
in field theory~\cite{chou,ww},
as the ones we are interested to occur in the 
neighbourhood of the advancing bubble walls. Indeed, what we
need is to compute the temporal evolution of classical order
parameters,
namely CP-violating particle currents, with definite initial
conditions.
In this respect, the ordinary equilibrium quantum field theory at
finite temperature may not be applied, since it mainly
deals with transition amplitudes in particle reactions.

It is well-known that self-energy
corrections at one- or two-loops to
the propagator modify the dispersion relations and
introduce nontrivial effects ({\it e.g.} damping) due to the imaginary
contributions to the self-energy~\cite{par}. This is because particles
propagating in the plasma experience interactions
with the surrounding particles of the thermal bath,
and the correct way to describe this phenomenon is to
substitute particles by quasiparticles and to adopt
dressed propagators. The self-energy takes the form
$\Sigma(k)={\rm Re}\: \Sigma(k)+i\:{\rm
Im}\:\Sigma(k)$. Due to the nonvanishing ${\rm
Im}\:\Sigma$,
the spectral function
$\rho({\bf k},k^0)$ acquires in the weak limit a finite width
\begin{eqnarray}
\Gamma(k)&=&-\frac{{\rm
Im}\:\Sigma({\bf k},\omega)}{2\:\omega(k)},\nonumber\\
\omega^2(k)&=&{\bf k}^2+m^2+{\rm Re}\:\Sigma({\bf k},\omega),
\end{eqnarray}
and is expressed by
\begin{equation}
\rho({\bf k},k^0)=i\:\left[\frac{1}{(k^0+i\:\varepsilon+ i\:
\Gamma)^2-\omega^2(k)}-
\frac{1}{(k^0-i\:\varepsilon-i\:\Gamma)^2-\omega^2(k)}\right].
\label{roigual}
\end{equation}
where $m$ indicates the tree-level mass.
It is easy to show that the free propagator is obtained when taking
the limit $\Gamma\rightarrow 0$.

The advantage of the method is that very
general formulae can be obtained for the temporal
evolution of CP-violating observables without any
particular assumption on the relative magnitude of the mean free
paths
and the thickness of the wall.
The standard thin wall and thick wall
limits are recovered in the limit $\Gamma L_{\omega}\rightarrow 0$
and $\infty$, respectively.
Moreover, because of the
presence of a finite damping rate, the Green's functions of the
particles involved in the processes are damped for times
$t\simgt \Gamma^{-1}$, reflecting the effects of the incoherent
nature of the plasma on CP-violating observables.

Following~\cite{newmethod1,newmethod2},
we are interested in the generation of charges which are
approximately conserved in the symmetric phase, so that they
can efficiently diffuse in front of the bubble where baryon number
violation is fast, and non-orthogonal to baryon number,
so that the generation of a non-zero baryon charge is energetically
favoured.
Charges with these characteristics
are the axial stop charge and the Higgsino charge,
which may be produced from the interactions of squarks and
charginos
and/or neutralinos with the bubble wall,
provided a source of CP-violation is present in these sectors.
CP-violating sources $\gamma_Q(z)$
(per unit volume and unit time) of a generic charge density
$J^0$ associated with  the current $J^\mu(z)$
and accumulated by the moving wall at a point
$z^\mu$ of the plasma can then be constructed from
$J^\mu(z)$~\cite{newmethod1,newmethod2}
\begin{equation}
\label{source}
\gamma_Q(z)=\partial_0 J^0(z)   .
\end{equation}
The definition of
the CP-violating source
$\gamma_Q(z)$ is  appropriate to describe the damping effects
originated by the
plasma interactions, but
does {\it not} incorporate any relaxation time scale
arising when diffusion and particle changing interactions are
included.
However, one can
leave aside diffusion and particle changing interactions
and account for them
independently in the rate
equations~\cite{newmethod1,newmethod2}.

\vspace{0.5cm}

{\bf 3.1} The right-handed stop current
$J^\mu_R$ associated to the right-handed stop $\widetilde{t}_R$
is given by,
\begin{equation}
J_R^\mu=i\left(\widetilde{t}^*_R\partial^\mu\widetilde{t}_R-
\partial^\mu\widetilde{t}^*_R \widetilde{t}_R\right).
\end{equation}
$\langle J_{R}^\mu(z)\rangle$ gets
contributions
from eight different one-loop triangle Feynman diagrams,
symbolically depicted in Fig.~1a.

In the calculation of $\langle J_R^0(z)\rangle$
at the lowest order in a ``Higgs insertion expansion'', and more precisely
in the integrand of $\int dx dy$ ($x$ and $y$ being the time components
of the four-vectors where the two insertions of the diagram in Fig.~1a
have been defined~\footnote{Hereafter the generic notation for the time
component of the four-vector $z^{\mu}$ will be adopted as $z$, {\it e.g.}
$z^{\mu}\equiv(z,\vec{z})$.}), the function
$H_2(x+z)H_1(y+z)-H_1(x+z)H_2(y+z)$ appears,
where $H_i$, $i = 1,2$, denote the two neutral Higgs components.
In order to deal with analytic expressions, we can work out
the thick wall limit and simplify the expressions obtained above
by performing a derivative expansion
\begin{equation}
\label{dd}
H_i(x+z)= \sum_{n=0}^{\infty}\frac{1}{n!}\; \frac{\partial^n}
{\partial z^n} H_i(z)\ x^n .
\end{equation}
The contribution to the current can then be expanded in a power series as:
\begin{eqnarray}
\label{expansion}
&&H_2(x+z)H_1(y+z)-H_1(x+z)H_2(y+z) \nonumber\\
&=&\sum_{n=0}^{\infty}
\frac{\partial^n}{\partial z^n}\left[H_1(z)\partial_z H_2(z)-H_2(z)
\partial_z H_1(z)\right]
f_n(x,y)
\end{eqnarray}
where the function $f_n$ is provided by the power expansion,
{\it e.g}. $f_0=x-y$, $f_1=(x^2-y^2)/2$, $\dots$  Notice that the term
with no
derivatives vanishes in the expansion (\ref{expansion}), {\it e.g.}
$H_2(z)H_1(z)-H_1(z)H_2(z)\equiv 0$, which means that the static
term in the derivative expansion (\ref{dd}) does not contribute
to the vacuum expectation value of the currents.
Furthermore, the approximation of neglecting terms in (\ref{dd})
with $n>1$ amounts to neglecting terms in (\ref{expansion}) with $n>0$.
Since for a smooth Higgs profile, the subsequent derivatives with
respect to the time coordinate are associated with higher
powers of $v_{\omega}/L_{\omega}$, and the integration over $x$ and $y$
of a higher order term in the expansion of the currents
leads to a higher power of $1/\Gamma_{\stop}$,
this is a good approximation for values of
$L_{\omega}\Gamma_{\stop}/v_{\omega} \gg 1$.
In other words, this expansion is valid only when the mean free path
$\tau_{\stop}\simeq
\Gamma^{-1}_{\stop}$ is smaller than the scale of variation of the Higgs
background determined by the wall thickness, $L_{\omega}$,
and the wall velocity $v_{\omega}$. An estimate
of the stop damping rate (in the low momentum limit)
is  $\Gamma_{\stop}\sim 10^{-1}\:T$, approximated using the calculation
made in
Ref.~\cite{EnqvistEV}. With such value, our derivative expansion is
perfectly justified since the wall thickness can span the range
$(10-100)/T$. 

Therefore, the currents are proportional to
the function (the coefficient of
$f_0$ in Eq. (\ref{expansion}) )
\begin{equation}
H_1(z)\partial_z H_2(z)-H_2(z) \partial_z H_1(z)
\equiv H^2(z) \partial_z\beta(z),
\label{current}
\end{equation}
which  should vanish smoothly for values of $z$ outside the
bubble wall. Here $H^2\equiv H_1^2+ H_2^2$.

Since the time variation of the Higgs fields is due to  the
expansion of the bubble wall through the thermal bath,
$\langle J_R^\mu(z)\rangle$
will be linear in $v_{\omega}$.
This result explicitly shows that we need out of equilibrium
conditions to generate $\langle J_R^\mu(z)\rangle$ and that we
have
to call for the CTP formalism to deal with time-dependent
phenomena.
To work out exactly  $\langle J_R^\mu(z)\rangle$ one should
know the exact form of the distribution functions
which, in ultimate analysis,
are provided by solving the Boltzmann equations.
However, any departure from thermal equilibrium distribution
functions
is caused at a given point by the passage of the wall and, therefore,
is  ${\cal O}(v_{\omega})$.  Since $\langle J_R^\mu(z)\rangle$ is
already linear in $v_{\omega}$,
working with thermal equilibrium distribution
functions
amounts to ignoring terms of higher order in
$v_{\omega}$~\cite{newmethod2},
which is as accurate as the bubble wall is moving slowly in
the plasma and we shall adopt this approximation from now on.

One can show that 
%\cite{bau2}
%
\begin{eqnarray}
\langle J_R^0(z)\rangle&=&\:h_t^2\:{\rm Im}
\left(A_t\mu\right)  \left[H_1(z) \partial_z H_2(z) -
H_2(z) \partial_z H_1(z)
\right] {\cal G}_R^Q
\label{avcorriente}
\end{eqnarray}
where
\begin{eqnarray}
{\cal G}_R^Q &=& \int_0^\infty dk \frac{k^2}{4 \pi^2 \Gamma_R  \;
\omega_Q \; \omega_R}   \nonumber\\
&\left[ \phantom{\frac{1}{2^2}} \right. &
 \left(1 + 2 {\rm Re}(n_Q) \right)
I_1(\omega_R,\Gamma_R,\omega_Q,\Gamma_Q)
+
\left(1 + 2 {\rm Re}(n_R) \right)
I_1(\omega_Q,\Gamma_Q,\omega_R,\Gamma_R) \nonumber\\
&-&
2 \left( {\rm Im}(n_R) +
{\rm Im}(n_Q) \right) I_2(\omega_R,\Gamma_R,
\omega_Q,\Gamma_Q) \left. \phantom{\frac{1}{2^2}} \right]
\end{eqnarray}
and $\omega^2_{R(L)}=k^2+m_{U(Q)}^2+\Pi_{U(Q)}$,
$n_{R(Q)} = 1/\left[\exp\left(\omega_{R(Q)}/T + i \Gamma_{R(Q)}/T
\right)- 1 \right]$, while $\Gamma_R\sim\Gamma_Q\sim\Gamma_{\stop}$.

The functions $I_1$ and $I_2$ are given by 
%\cite{bau2}
\begin{eqnarray}
I_1(a,b,c,d) = \frac{r_1}{\left(r_1^2 + 1 \right)
\left[(a+c)^2 + (b+d)^2 \right]}
+ \frac{r_2}{\left(r_2^2 + 1 \right)\left[(a-c)^2 + (b+d)^2 \right]},
\end{eqnarray}
\begin{eqnarray}
I_2(a,b,c,d) = \frac{r_1^2 - 1}{2 \left(r_1^2 + 1 \right)
\left[(a+c)^2 + (b+d)^2 \right]} +
\frac{r_2^2 - 1}{2 \left(r_2^2 + 1 \right)
\left[(a-c)^2 + (b+d)^2 \right]},
\end{eqnarray}
where $r_1 = (a+c)/(b+d)$ and $r_2 = (a-c)/(b+d)$.

\vspace{0.5cm}
{\bf 3.2} Similarly, the Higgsino current associated with  neutral
and charged Higgsinos can be written
as
\be
\label{corhiggs}
J^{\mu}_{\widetilde{H}}=\overline{\widetilde{H}}\gamma^\mu \widetilde{H}
\ee
where $\widetilde{H}$ is the Dirac spinor
\be
\label{Dirac}
\widetilde{H}=\left(
\begin{array}{c}
\widetilde{H}_2 \\
\overline{\widetilde{H}}_1
\end{array}
\right)
\ee
and $\widetilde{H_2}=\widetilde{H}_2^0$ ($\widetilde{H}_2^+$),
$\widetilde{H_1}=\widetilde{H}_1^0$ ($\widetilde{H}_1^-$) for
neutral (charged) Higgsinos.

The Higgsino current (\ref{corhiggs})
gets its contribution from the typical triangle diagram of
Fig.~1b and may be computed along the same lines described for
the
axial stop number in the previous subsection.
Analogously to the case of right-handed stops, the dispersion
relations of charginos and neutralinos are changed by high
temperature
corrections~\cite{weldon}. Even though fermionic dispersion
relations
are highly nontrivial, relatively simple expressions
for the fermionic spectral functions may be given in the limit
in which the damping rate is smaller than the typical self-energy
of the fermionic excitation (this limit is certainly
satisfied in our case)~\cite{henning}. For instance, the spectral
function of Higgsinos $\widetilde{H}$ (and analogously for
gauginos $\widetilde{W},\widetilde{B}$) may be written as
\begin{eqnarray}
\rho_{\widetilde{H}}({\bf k},k^0)& = & i
\left(\not k + m_{\widetilde{H}}\right) \\
&&
\left[\frac{1}{(k^0+i\:\varepsilon+ i\:
\Gamma_{\widetilde{H}})^2-\omega_{\widetilde{H}}^2(k)}-
\frac{1}{(k^0-i\:\varepsilon-i\:\Gamma_{\widetilde{H}})^2
-\omega_{\widetilde{H}}^2(k)}\right],\nonumber
\label{rofermion}
\end{eqnarray}
where $\omega_{\widetilde{H}}^2(k)={\bf k}^2 +
m_{\widetilde{H}}^2(T)$ and $m_{\widetilde{H}}^2(T)$
is the Higgsino effective plasma squared mass in the thermal bath
which
may be well approximated by its value in the present vacuum,
$m_{\widetilde{H}}^2(T)\simeq |\mu|^2$. Similarly, $|\mu|$
should
be replaced by $M_2$ for $\rho_{\widetilde{W}}({\bf k},k^0)$,
and by $M_1$ for $\rho_{\widetilde{B}}({\bf k},k^0)$.

In analogy to the stop case, in the region of parameters
defined above, we can perform
a Higgs insertion expansion of the CP-violating current.
The vacuum expectation value of the (zero component of the)
Higgsino current is then found to be 
%\cite{bau2}
\begin{equation}
\langle J_{\widetilde{H}}^0(z)\rangle =
{\rm Im}(\mu)\: \left[H_1(z) \partial_z H_2(z) - H_2(z)
\partial_z H_1(z) \phantom{1^2} \right]
\left[ 3 M_2 \; g_2^2 \; {\cal G}^{\widetilde{W}}_{\widetilde{H}}
 +       M_1 \; g_1^2 \; {\cal G}^{\widetilde{B}}_{\widetilde{H}}
\right],
\end{equation}
where
\begin{eqnarray}
{\cal G}^{\widetilde{W}}_{\widetilde{H}} & = & \int_0^\infty dk
\frac{k^2}
{4 \pi^2 \Gamma_{\widetilde{H}}
\omega_{\widetilde{H}} \omega_{\widetilde{W}}} \nonumber\\
&\left[ \phantom{\frac{1}{2^2}} \right.&
 \left(1 - 2 {\rm Re}(n_{\widetilde{W}}) \right)
I_1(\omega_{\widetilde{H}},\Gamma_{\widetilde{H}},
\omega_{\widetilde{W}},\Gamma_{\widetilde{W}})+
\left(1 - 2 {\rm Re}(n_{\widetilde{H}}) \right)
I_1(\omega_{\widetilde{W}},
\Gamma_{\widetilde{W}},\omega_{\widetilde{H}},
\Gamma_{\widetilde{H}}) \nonumber\\
&+&
2 \left( {\rm Im}(n_{\widetilde{H}}) +
{\rm Im}(n_{\widetilde{W}}) \right)
I_2(\omega_{\widetilde{H}},
\Gamma_{\widetilde{H}},
\omega_{\widetilde{W}},\Gamma_{\widetilde{W}})
\left.\phantom{\frac{1}{2^2}} \right]
\nonumber\\
\end{eqnarray}
and $\omega^2_{\widetilde{H}(\widetilde{W})}=k^2+ |\mu|^2
(M_2^2)$ while $n_{\widetilde{H}(\widetilde{W})} =
1/\left[\exp\left(\omega_{\widetilde{H}(\widetilde{W})}/T
+ i \Gamma_{\widetilde{H}(\widetilde{W})}/T \right)
+ 1 \right]$.
The damping rate of charged and neutral
Higgsinos is
dominated by weak interaction and we take
$\Gamma_ {\widetilde{H}} \simeq \Gamma_{\widetilde{W}}$
to be of order of $ 5\times 10^{-2} T$.
The Bino contribution may be obtained from the above
expressions by replacing $M_2$ by $M_1$.

\vspace{1cm}
\noindent
{\bf 4.} We can now solve the set of coupled differential
equations describing the effects of diffusion, particle number
changing reactions and CP-violating source terms.
We will closely follow the approach taken in
Ref.~\cite{newmethod2} where the
reader is referred to for more details.
If the system is near thermal equilibrium and particles interact weakly,
the particle number densities $n_i$ may be expressed as
$n_i=k_i\mu_iT^2/6$ where
$\mu_i$ is the local chemical potential, and $k_i$ are statistical
factors of order of 2 (1) for light bosons (fermions) in thermal
equilibrium, and Boltzmann suppressed for particles heavier than $T$.

The particle densities we
need to include are the left-handed top
doublet $q_L\equiv(t_L+b_L)$,
the right-handed top quark $t_R$, the Higgs particle
$h\equiv(H_1^0, H_2^0, H_1^-, H_2^+)$, and the superpartners
$\widetilde{q}_L$, $\widetilde{t}_R$ and $\widetilde{h}$.
The interactions able to change the particle numbers are the top
Yukawa interaction with rate
$\Gamma_t$, the top quark mass interaction with rate $\Gamma_m$,
the Higgs self-interactions in the broken phase
with rate $\Gamma_{\cal H}$, the strong sphaleron interactions
with rate $\Gamma_{{\rm ss}}$,
the weak anomalous interactions with rate $\Gamma_{\rm ws}$
and the gauge interactions.
We shall assume that the supergauge interactions are in
equilibrium.
Under these assumptions the system may be described by the
densities
${\cal Q} = q_L + \widetilde{q}_L$,
${\cal {\cal T}}=t_R+\widetilde{t}_R$ and ${\cal H}=h+\widetilde{h}$.
CP-violating interactions with the advancing bubble wall produce
source terms $\gamma_{\widetilde{H}}=\partial_z\langle
J_{\widetilde{H}}^0(z)\rangle$ for Higgsinos and
$\gamma_R=\partial_z \langle J_{R}^0(z)\rangle$
for right-handed stops, which tend to push the system out of
equilibrium.
Ignoring the curvature of the bubble wall, any quantity becomes a
function of the coordinate
${\bf z}=z_3+v_{\omega}z$, the coordinate normal
to the wall surface, where we assume the bubble wall is moving along the
$z_3$-axis.

Assuming that the rates $\Gamma_t$ and $\Gamma_{{\rm ss}}$ are
fast so that ${\cal Q}/k_q-
{\cal H}/k_{\cal H}-{\cal T}/k_{\cal T}={\cal O}(1/\Gamma_t)$ and
$2{\cal Q}/k_q-{\cal T}/k_{\cal T}+
9({\cal Q}+{\cal T})/k_b={\cal O}(1/\Gamma_{{\rm ss}})$,
one can find the equation governing the Higgs density
\begin{equation}
\label{equation}
v_{\omega}{\cal H}^\prime-\overline{D}
{\cal H}^{\prime\prime}+\overline{\Gamma}{\cal H}-
\widetilde{\gamma}=0,
\end{equation}
where the derivatives are now with respect to ${\bf z}$,
$\overline{D}$ is the effective diffusion constant,
$\widetilde{\gamma}$ is an effective source term
in the frame of the
bubble wall and $\overline{\Gamma}$ is the effective decay
constant~\cite{newmethod2}.
An analytical solution to Eq.~(\ref{equation}) satisfying the
boundary conditions ${\cal H}(\pm\infty)=0$ may be found in the
symmetric
phase (defined by ${\bf z}<0$) using a ${\bf z}$-independent
effective diffusion constant (in agreement with our approximation
in Eq.~(\ref{dd})) and a step function for the effective decay rate
$\overline{\Gamma}= \widetilde{\Gamma} \theta({\bf z})$. A more realistic
form of $\overline{\Gamma}$ would interpolate smoothly between the
symmetric and the broken phase values. We have checked, however,
that the result is insensitive to the specific position of the
step function  inside the bubble wall. 
The values of $\overline{D}$ and $\overline{\Gamma}$ in
(\ref{equation}) of course depend on the particular
values of supersymmetric parameters. For the
considered range we typically find $\overline{D}\sim 0.8\ {\rm GeV}^{-1}$,
$\overline{\Gamma}\sim 1.7$ GeV.

The dependence on $\widetilde{\gamma}$
is much more subtle, and a good
approximation for the Higgs profiles is necessary in order
to obtain a reliable approximation to the currents. As seen in
Eq.~(\ref{current}), the current is proportional to the
variation of the ratio of vacuum expectation values along
the wall. The Higgs profiles along the wall are likely to
follow the path of minimal energy connecting the false and
the true vacua in the Higgs potential. Close to the symmetric
phase, and at the temperature $T_0$, defined as the one at
which the curvature at the origin vanishes,
the ratio of vacuum expectation values is
given by  $\tan^2 \beta({\bf z}=0) = m_1^2(T_0)/m_2^2(T_0)$,
a quantity which may be computed numerically from the
Higgs finite temperature effective potential as
$m_i^2(T)=1/2 \ \left(\partial^2 V_{\rm eff}/\partial H_i^2\right)
|_{H_1=H_2=0},\ (i=1,2)$. Hence, the
variation of $\beta$ is given by
\begin{equation}
\Delta\beta = \beta(T_0) - \arctan(m_1(T_0)/m_2(T_0))
\label{deltabeta}
\end{equation}
This quantity tends to zero for large values of $m_A$,
and takes small values, of order $0.015$ for values of
$m_A = 150$--200 GeV, giving rise to a further suppression of the
final baryon number.

One would be tempted to adopt  a linear ansatz already for
the function (\ref{current}) in the current,
as done in \cite{newmethod2}. However, this approximation has the drawback that
it  may not be  obtained starting from smooth Higgs profiles and can
therefore be viewed as unphysical. Also, since
the source in the diffusion equation depends on the
derivatives of the current, linear walls (or, generally, all non-differentiable
wall shapes) lead to the appearence of $\delta$-functions in the CP-violating
source.
A more realistic option is to specify the shapes of the Higgs fields and
$\beta$ and to consequently compute the source. Realistic shapes for
$H({\bf z})$ (and probably for $\beta({\bf z})$)
would be given by functional kinks along the bubble wall.
This provides a smooth function for the source $\widetilde{\gamma}$. Moreover,
this way of proceeding leads to
a strong suppression with respect to  the result that one would have obtained
assuming   a linear approximation for the quantity
$H_1(z)\partial_z H_2(z)-H_2(z) \partial_z H_1(z)$. This is due to the fact
that, since $H_1(z)\partial_z H_2(z)-H_2(z) \partial_z H_1(z)$ goes to zero
for ${\bf z}\rightarrow \pm \infty$, the source $\widetilde{\gamma}$ must
change sign inside the bubble wall, giving rise to some cancellation for the
final expression of the baryon number, see Eqs.~(29) and (32).

We consider here the following semirealistic approximation:
\begin{eqnarray}
H(\bf{z}) & = & \frac{v(L_{\omega})}{2}
\left[ 1 -
\cos\left(\frac{{\bf{z}} \pi}{ L_{\omega}} \right)
\right]
\left[ \theta({\bf{z}}) - \theta({\bf{z}}-L_{\omega}) \right]
+ v(L_{\omega}) \theta({\bf{z}}-L_{\omega})
\nonumber\\
\beta(\bf{z}) & = & \Delta\beta \left[ 1 -
\cos\left(\frac{{\bf{z}} \pi}{L_{\omega}}\right)
\right]
\left[ \theta({\bf{z}}) - \theta({\bf{z}}-L_{\omega}) \right]
+ \beta({\bf{z}}=0) + \Delta\beta \theta({\bf{z}}-L_{\omega})
\label{profiles}
\end{eqnarray}
which leads to a current vanishing smoothly at ${\bf{z}}=0$ and
${\bf{z}}=L_{\omega}$~\footnote{We have checked that numerical results
using the ansatz (\ref{profiles}) differ by less than 10\% from those
obtained by using functional kinks. This leads us to the belief
that our results are robust even in the absence of a numerical
calculation of bubble solutions in the MSSM.}.

For ${\bf z} < 0$ one obtains,
\begin{equation}
\label{higgs1}
{\cal H}({\bf z})={\cal A}\:{\rm e}^{{\bf z}v_{\omega}/\overline{D}},
\end{equation}
and for ${\bf z} >0$,
\begin{eqnarray}
\label{higgs3}
{\cal H}({\bf z}) & = & \left( {\cal B}_{+} -
\frac{1}{\overline{D}(\lambda_+  - \lambda_-)}
\int_0^{{\bf z}} du \widetilde \gamma(u) e^{-\lambda_+ u} \right)
e^{\lambda_{+} {\bf z}}
\nonumber\\
&+& \left( {\cal B}_{-} -
\frac{1}{\overline{D}(\lambda_-  - \lambda_+)}
\int_0^{\bf{z}} du \widetilde \gamma(u) e^{-\lambda_- u} \right)
e^{\lambda_{-} {\bf z}}.
\end{eqnarray}
where
\begin{equation}
\lambda_{\pm} = \frac{ v_{\omega} \pm
\sqrt{v_{\omega}^2 + 4 \widetilde{\Gamma}
\overline{D}}}{2 \overline{D}},
\end{equation}
and $\widetilde \gamma({\bf z}) = v_{\omega} \partial_{{\bf z}}
J^0({\bf z}) f(k_i)$,
$J_0$ being the total CP-violating current resulting
from the sum of the right-handed
stop and Higgsino contributions and
$f(k_i)$  a coefficient depending on the number of
degrees of freedom present in
the thermal bath and related to the definition of the
effective source~\cite{newmethod2}.
Imposing the continuity of ${\cal{H}}$ and
${\cal{H}}'$ at the boundaries, we find
\begin{equation}
\label{higgs2}
{\cal A}= {\cal B}_{+}\left(1-\frac{\lambda_-}{\lambda_+}\right)=
{\cal B}_{-}\left(\frac{\lambda_+}{\lambda_-}-1\right)=
\frac{1}{\overline{D} \; \lambda_{+}} \int_0^{\infty} du\;
\widetilde \gamma(u)
e^{-\lambda_+ u}.
\end{equation}
From the form of the above equations one can see that CP-violating
densities are non zero for a time $t\sim \overline{D}/ v_{\omega}^2$
and the assumptions leading to the analytical
form of ${\cal H}({\bf z})$ are valid
provided $\Gamma_t,\Gamma_{{\rm ss}}\gg
v_{\omega}^2/\overline{D}$.

The equation governing
the baryon asymmetry $n_B$ is given by~\cite{newmethod2}
\begin{equation}
\label{bau}
D_q n_B^{\prime\prime}-v_{\omega} n_B^\prime-
\theta(-{\bf z})n_f\Gamma_{{\rm ws}}n_L=0,
\end{equation}
where $\Gamma_{{\rm ws}}=6\kappa\alpha_w^4T$ is the
weak sphaleron
rate ($\kappa\simeq 1$)~\cite{SphalRate}\footnote{The
correct value of $\kappa$ is at present the subject
of debate; see, for instance, Ref.~\cite{ASY}.},
and $n_L$ is the total number density of
left-handed weak doublet fermions, $n_f=3$ is the number of
families and
we have assumed that the baryon asymmetry gets
produced only in the symmetric phase.
Expressing $n_L({\bf z})$ in terms of the Higgs number density
\begin{equation}
n_L=\frac{9k_q k_{\cal T}-8k_b k_{\cal T}
-5 k_b k_q}{k_{\cal H}(k_b+9 k_q+9 k_{\cal T})}\:{\cal H}
\end{equation}
and making use of Eqs.~(\ref{higgs1})-(\ref{bau}), we find that
\begin{equation}
\frac{n_B}{s}=-g(k_i)\frac{{\cal A}\overline{D}\Gamma_{{\rm ws}}}
{v_{\omega}^2 s},
\end{equation}
where $s=2\pi^2 g_{*s}T^3/45$ is the entropy density ($g_{*s}$
being
the effective number of relativistic degrees of freedom) and
$g(k_i)$
is a numerical coefficient depending upon the light degrees of
freedom present in the thermal bath.

\vspace{1cm}
\noindent
{\bf 5.} The expression for the baryon asymmetry derived in the last
section
assumes that the order parameter $v(T_c)/T_c \geq 1$, in order to
assure that anomalous processes are sufficiently suppressed in the
broken phase. As we already discussed, this demands low values
of the right-handed stop effective plasma mass at the critical
temperature and low values of the mixing mass parameter
$\widetilde{A}_t$.
We will now present numerical results based on the
previous calculations.
Since $\Delta\beta$, Eq.~(\ref{deltabeta}),
denotes the variation of the ratio of vacuum
expectation values of the Higgs fields along the bubble wall, it will
be negligible unless  the CP-odd Higgs mass takes values of
the order of the critical temperature. Indeed, for large
values of $m_A$, $\Delta\beta \sim 1/m_A^2$.

Values of the CP-odd Higgs mass $m_A \simlt 200$ GeV
are associated with a weaker first order phase transition. Fig.~2
shows the behaviour of the order parameter $v/T$ in the
$m_A$-$\tan\beta$ plane, for $\widetilde{A}_t = 0$, $m_Q = 500$
GeV and values of $\widetilde{m}_U$ close to its upper bound,
Eq.~(\ref{colorbound}). In order to correctly interpret the results
of Fig.~2 one should remember that the Higgs mass bounds are somewhat
weaker for values of $m_A < 150$ GeV. However, even for values
of $m_A$ of order 80 GeV, in the low $\tan\beta$ regime the lower
bound on the Higgs mass is of order 60 GeV. Hence, it follows
from Fig.~2 that, to obtain a sufficiently strong first order
phase transition, $v/T \simgt 1$, the CP-odd Higgs mass
$m_A \simgt 150$ GeV~\footnote{Recent analyses~\cite{FL} of the
effective three-dimensional theory of the MSSM essentially
confirm the early results of Ref.~\cite{mariano2}, while a
similar analysis for the light stop scenario is still lacking.}.
In our analysis, we shall fix
$m_A = 200$ GeV, which is well inside the acceptable range for
this parameter. As anticipated above, this choice leads to values
of $\Delta\beta \simeq 0.015$.

Since for large values of $m_Q$
the stop contribution to the baryon asymmetry is strongly
suppressed compared to the chargino and neutralino ones,
the  numerical values of the baryon asymmetry depend linearly
on the phase of the Higgsino mass parameter. In general, the
numerical values obtained following the above procedure are
too low to explain the observed baryon asymmetry unless this
phase is of order one. This conclusion may only be avoided for
certain specific values of the gaugino and Higgsino mass parameters.
In order to make this result explicit, in Fig.~3
we show the value of
this phase necessary to obtain a value $n_B/s \simeq 4 \times 10^{-11}$
in the $\mu$-$M_2$ plane. The wall velocity is
taken to be $v_{\omega} = 0.1$, while the bubble wall width is
taken to be $L_{\omega} = 25/T$, and, for simplicity, we consider
$M_1 = M_2$~\footnote{We have checked that, as expected, the final
results do not depend much on the value of $M_1$. In particular
$M_1$ can be chosen such that the lightest neutralino is lighter
than the light stop.}. Our results, are, however, quite  insensitive to
the specific choice of $v_{\omega}$ and $L_{\omega}$.
Due to the fact that, as already stated,
stop contributions to the baryon asymmetry are negligible
as compared to the chargino/neutralino ones, we have taken
$| \sin(\phi_A+\phi_\mu)| =0$ in Fig.~3.

Values of the phases lower than 0.1 are only consistent with
the observed baryon asymmetry for values of $|\mu|$ of
order of the gaugino mass parameters. This is due to a large
enhancement of the computed baryon asymmetry for these values
of the parameters. 
Since the Higgs background in the Feynman diagrams of Fig.~1b is
carrying a very low momentum (of order of the inverse of the bubble wall
width $L_\omega$), this resonant behaviour is associated with the
possibility of absortion (or emission) of Higgs quanta by the
propagating particles. For momenta of order of the critical
temperature, this can only take place when the Higgsinos and 
gauginos are nearly degenerate in mass, $\mu\sim M_2$.
By using the Uncertainty Principle, it is easy to understand that the
width of this resonance is expected
to be proportional to the decay width of the -inos, which is what comes
out from our explicit computation. In the case of the stops, instead,
since $m^2_Q\gg \widetilde{m}_U^2$, the same phenomenon can only happen for
momenta larger than $m_Q$. Such configurations are
exponentially suppressed and do not give any relevant
contribution to the one loop diagram of Fig.~1a.
The resonant behaviour is certainly enhanced by our
approximations. Indeed, since
we are working at first order in the Higgs insertions, the
relevant masses propagating in the loops are given by their
expressions in the symmetric phase. Hence, strong degeneracies
are possible when $\mu$ is close to $M_2$. This degeneracy,
however, will be broken by further Higgs insertions.

Therefore, taking the results from Fig.~3 at face value,
one would find an absolute lower
bound on $|\sin(\phi_\mu)|\simgt  10^{-2}$.
However, given all uncertainties in the calculation, including
those associated to the
different damping rates for squarks, Higgsinos and gauginos,
and the sphaleron parameters, in particular
$\kappa=\Gamma_{{\rm ws}}/
[6\alpha_w^4 T]$~\footnote{
We have adopted in the calculation
the value $\kappa=1$. The final result scales with $\kappa$.}, we
will prefer to consider, as a conservative estimate, the previous
bound on $|\sin(\phi_\mu)|$ accurate only up to about one order
of magnitude. If the phases of the parameters $A_t$ and $\mu$ 
are  ${\cal O}(1)$, CP-violation will
be induced in the Higgs sector and the Higgs VEVs will become complex.
However, it is easy to prove that, whenever the tree level CP-odd
mass is larger than the vacuum expectation values and the 
trilinear mass parameters $|A_t|$ and $|\mu|$ take values lower
than $m_Q$, the induced phase of the product of vacuum expectation
values will be much lower than 1, hence our results will be only
weakly modified by this effect.

\vspace{1cm}\noindent
{\bf 6.} In conclusion, we made a self-consistent computation of
the baryon asymmetry generated at the electroweak phase transition
in the Minimal Supersymmetric Standard Model. We  proved that
baryon asymmetry is mainly generated by CP-violating Higgsino
currents, provided that Higgsinos and gauginos are not much
heavier than the electroweak critical temperature ($T_c\sim 100$
GeV), and the phase $\phi_\mu$ is not much smaller than 0.1.
Observe that these relatively large values of the phases
are only consistent with the constraints from the
electric dipole moment of the neutron if the squarks of the
first and second generation have masses of the order of 
a few TeV~\cite{moremin}.
On the other hand, the generated baryon asymmetry is not washed
out inside the bubbles, after the phase transition, provided
that the lightest stop is lighter than the top quark (the
so-called light stop scenario), the pseudoscalar Higgs boson is
heavier than $\sim 150$ GeV and the lightest Higgs boson is
lighter than $\sim 80$ GeV.

The most direct experimental way of testing this scenario is
through the search for the lightest Higgs at LEP2.
If the Higgs is found, the second
test will come from the search for the lightest stop at the Tevatron
collider (the stop mass is typically too large for the stop to be seen
at LEP). If both particles are found, the last crucial test will
come from $B$ physics, more specifically,
in relation to the CP-violating effects.
Depending on their gaugino-Higgsino composition,
light neutralinos or charginos, at the reach of LEP2,
although not essential, would further restrict the allowed
parameter space consistent with electroweak baryogenesis.

Finally, let us expand on the Higgs searches. If the next run of LEP  at
$\sqrt{s}$ = 186 GeV collects  the
planned  luminosity, it will be able to search for Higgs masses
up to about 75 GeV~\cite{LEPRep}. If no Higgs signal is found, this will
pose additional constraints on the present scenario. In particular, in
order to get still a sufficiently strong first order phase transition
it will demand $\widetilde{A}_t/m_Q \simeq 0$. Moreover, the selected
parameter space leads
to values of the branching ratio ${\rm BR}(b\rightarrow s\gamma)$
different from the Standard Model case. Although the exact value
of this branching ratio depends strongly on the value of the $\mu$
and $A_t$ parameters, the typical difference with respect to the
Standard
Model prediction is of the order of the present experimental
sensitivity
and hence in principle testable in the near future. Indeed, for the
typical spectrum considered here, due to the light charged Higgs,
the
branching ratio ${\rm BR}(b \rightarrow s \gamma)$ is somewhat
higher than in the SM case, unless
negative values of $A_t\mu$  are
present.

\vskip1cm
\centerline{\large\bf Acknowledgements}
\vskip 0.2cm

We would like to thank R. Garisto, P. Huet, A. Nelson
and M. Shaposhnikov for useful
discussions. I. Vilja is grateful for the kind hospitality
offered
during his visit at the Fermilab Astrophysics Center
where part of this work was done. A. Riotto would like to
thank R. Kolb
whose never-ending skepticism about the idea of
electroweak baryogenesis spurred our efforts.
C. Wagner would like to thank the members of the
Theory Group at Fermilab and at the University of
Chicago, where part of this work has been done,
for their kind hospitality. The Aspen Center for
Physics and the Benasque Center for Physics, where
part of this work has been done, are kindly
acknowledged by M. Carena and C. Wagner, and M.
Quir\'os, respectively. A. Riotto is
supported by the DOE and NASA under Grant NAG5--2788.

\newpage
%%%%%%%%%%%%%%%%%%%%%%%%figure%%%%%%%%%%%%%%%%%%
\begin{figure}
%\psdraft
\centerline{
\psfig{figure=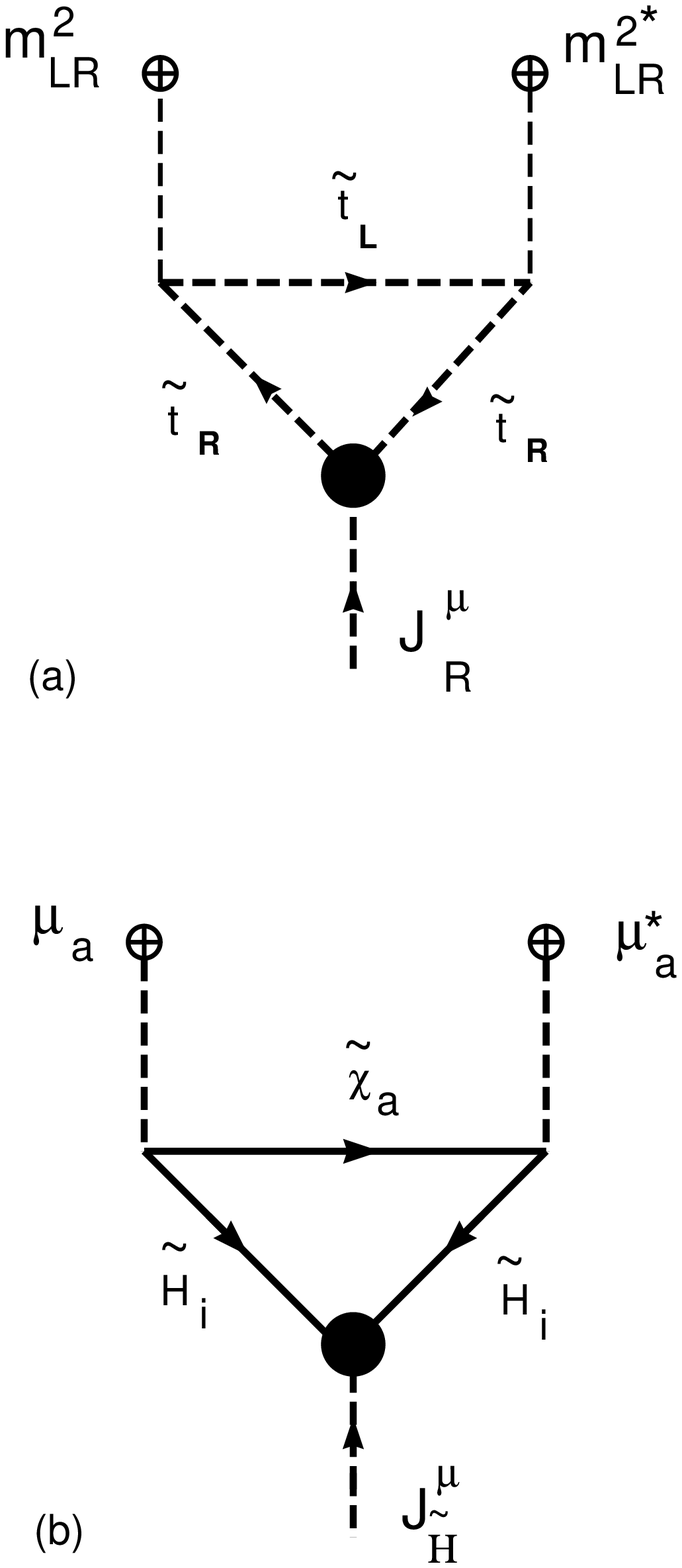,width=12cm,height=15.0cm}}
\caption{a) One-loop Feynman diagrams contributing to
$\langle J_R^{\mu}(z)\rangle$. Here $m_{LR}^2$ indicates the combination
$h_t\left(A_t H_2-\mu^{\star}H_1\right)$. b) One-loop diagrams contributing
to $\langle J^0_{\widetilde{H}}(z)\rangle$. Here $\mu_a$,
$a=1,\cdots,4$ describes the interactions of Higgsinos with gauginos
$\widetilde{W}_a$, $a=1,\cdots,3$ and $\widetilde{B}$, $a=4$.
More precisely, $\mu_a=g_a
\left[ H_1 P_L+\frac{\mu}{|\mu|} H_2 P_R\right]$ where
$P_{L(R)}=\frac{1}{2}\left(1\mp \gamma_5\right)$, $g_a=g_2$,
$a=1,\cdots,3$,  and $g_a=g_1$ for $a=4$.}
\label{feynman}
\end{figure}
%%%%%%%%%%%%%%%%%%%%%%%%figure%%%%%%%%%%%%%%%%%%
%%%%%%%%%%%%%%%%%%%%%%%%figure%%%%%%%%%%%%%%%%%%
%%%%%%
\begin{figure}
%\psdraft
\centerline{
\psfig{figure=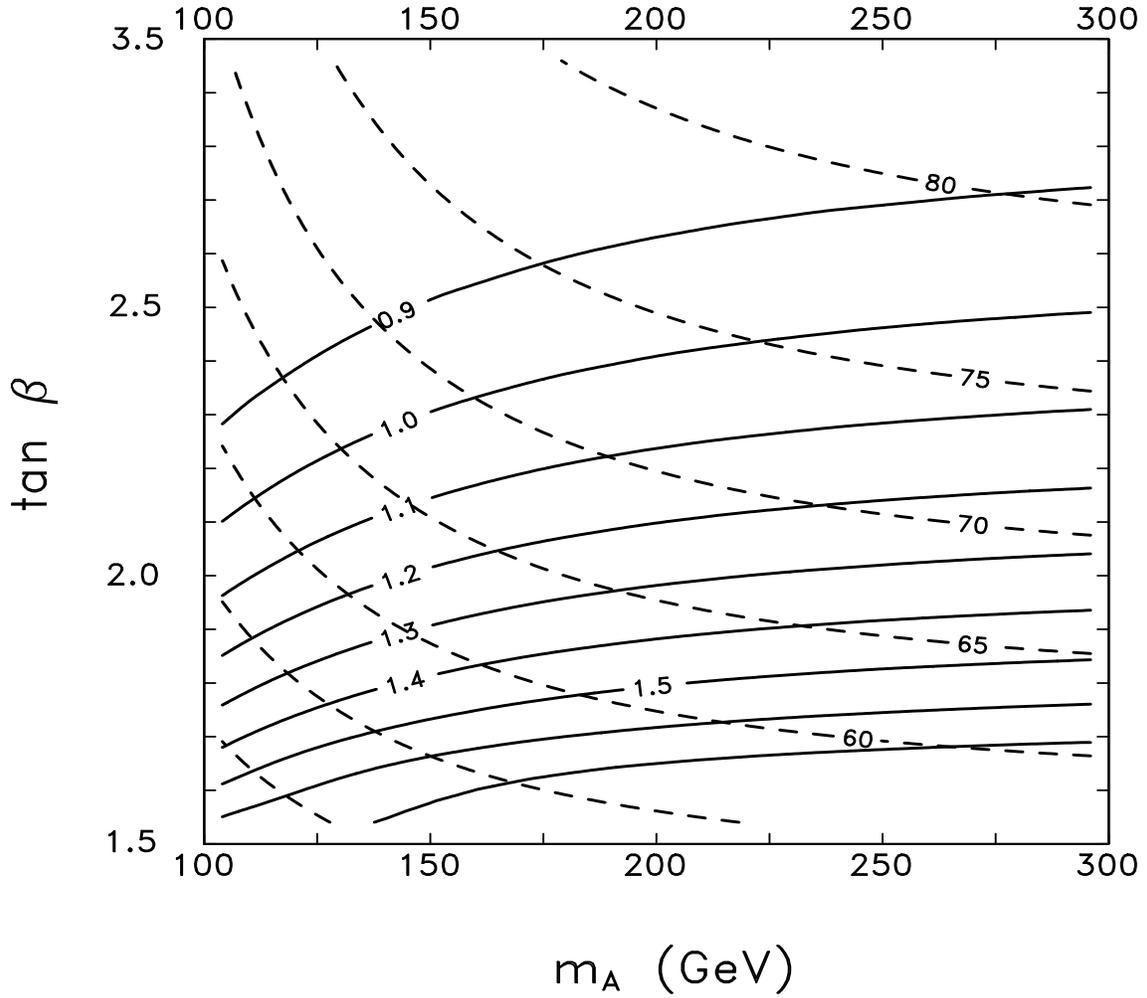,width=8cm,height=10.0cm,bbllx=8.cm,bblly=.cm,bburx=17.5cm,bbury=13cm}}
\caption{Contour plots of constant values of $v(T_c)/T_c$ (solid
lines) and $m_H$ in GeV (dashed lines) in the plane
$(m_A,\tan\beta)$. We have fixed $m_t=175$ GeV and the values
of sypersymmetric parameters: $m_Q=500$ GeV, $m_U=m_U^{\rm crit}$
fixed by the charge and color breaking constraint, and
$A_t=\mu/\tan\beta$.}
\label{f1}
\end{figure}
%%%%%%%%%%%%%%%%%%%%%%%%figure%%%%%%%%%%%%%%%%%%
%%%%%%
\newpage
%%%%%%%%%%%%%%%%%%%%%%%%figure%%%%%%%%%%%%%%%%%%
%%%%%%
\begin{figure}
%\psdraft
\centerline{
\psfig{figure=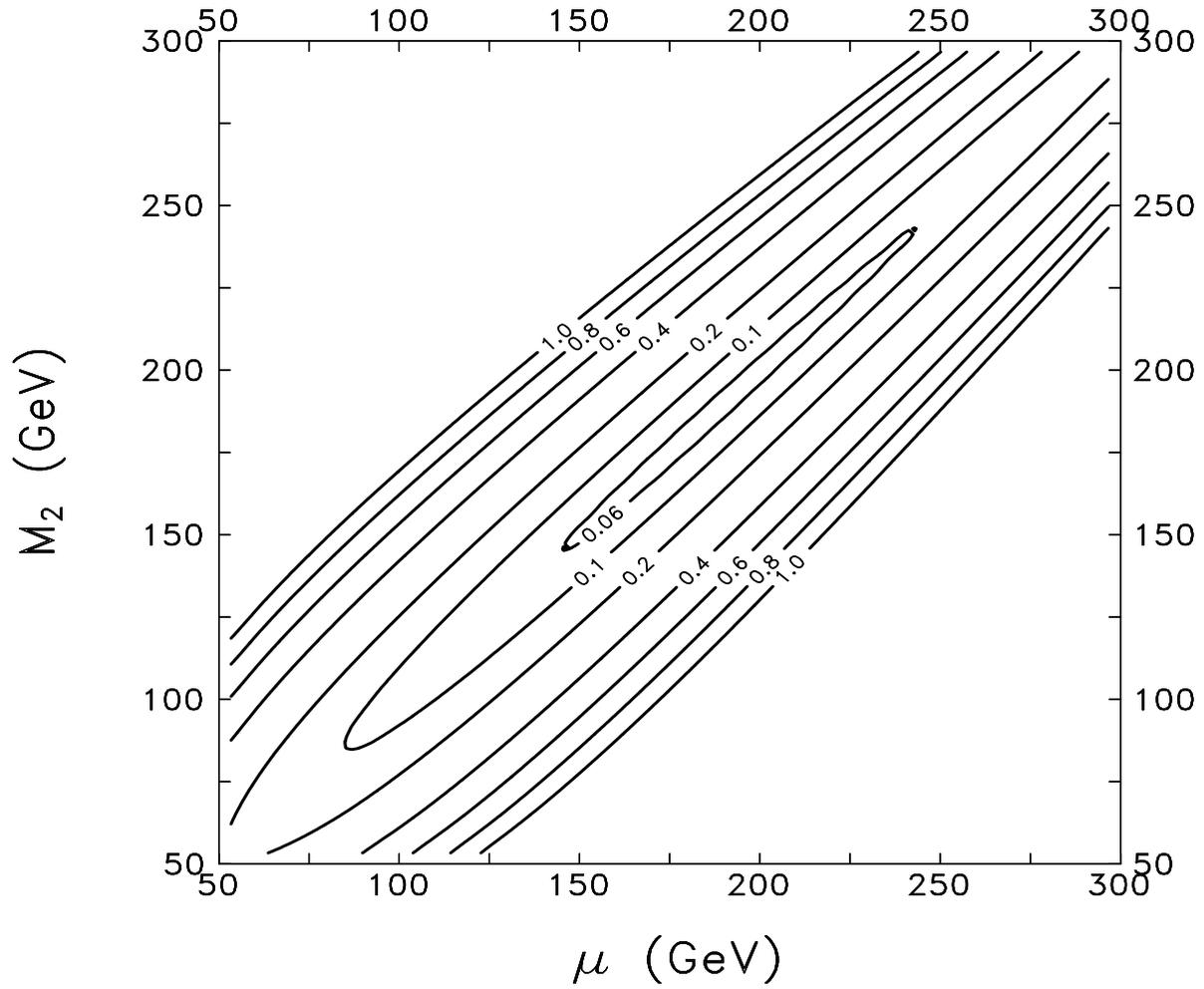,width=8cm,height=10.0cm,bbllx=5.8cm,bblly=.cm,bburx=15.3cm,bbury=13cm}}
\caption{Contour plot of $|\sin \phi_{\mu}|$
in the plane ($\mu,M_2$) for fixed $n_B/s = 4 \times 10^{-11}$ and
$v_{\omega}=0.1$, $L_{\omega}=25/T$, $m_Q=500$ GeV,
$m_U=m_U^{\rm crit}$, $\tan\beta=2$ and $A_t=\mu/\tan\beta$.}
\label{f2}
\end{figure}
%%%%%%%%%%%%%%%%%%%%%%%%figure%%%%%%%%%%%%%%%%%%
%%%%%%
\end{document}